# A reduced micromorphic model for multiscale materials and its applications in wave propagation

Mohamed Shaat*

*Engineering and Manufacturing Technologies Department, DACC, New Mexico State University, Las Cruces, NM 88003, USA*

*Mechanical Engineering Department, Zagazig University, Zagazig 44511, Egypt*

## Abstract

In this study, a reduced micromorphic model for multiscale materials is developed. In the context of this model, multiscale materials are modeled with deformable microstructures. The deformation energy is formed depending on microstrain and macroscopic strain residual fields. The constitutive equations according to the reduced micromorphic model only depend on eight material coefficients for linear elastic materials. These material coefficients are related to the material micro/macro-stifnesses and the material's microstructural features. The wave dispersions in multiscale materials are then derived according to the reduced micromorphic model. It is revealed that this model can reflect nine dispersion curves (three acoustic modes and six optics) for a two-scale material. To demonstrate the effectiveness of the proposed model, the wave propagation characteristics, the band structure, and the absolute bandgap features of phononic materials are investigated. It is demonstrated that the reduced micromorphic model can effectively reflect the increase in the bandgap width with the increase in the filling factor in a composite phononic material with square lattices.

**Keywords:** multiscale materials; bandgap; periodic; microstructure; micromorphic; phononic materials.

## 1. Introduction

Multiscale materials can be defined as materials exhibit different physical phenomena at different length and/or time scales. In order to capture two or more of these scale-dependent phenomena, a multiscale modeling approach should be utilized. In the context of this approach, length and time scales are defined to relate the different material's phenomena to their corresponding scales. Recently, multiscale modeling has intensively grown to cover different approaches and models that can reveal the mechanical, electrical, and magnetic properties of multiscale materials.

---

∗Corresponding author: *E-mail address:* shaat@nmsu.edu; shaatscience@yahoo.com (M. Shaat).
Tel.: +15756215929

In fact, materials that can be considered as multiscale materials are numerous. For example, materials such as polycrystalline materials, composites, superlattices, framework and molecular crystals, and metamaterials are multiscale materials. As a general classification, multiscale materials can be considered with periodic or non-periodic microstructures. Depending on the periodicity of the material microstructure and/or the scale at which the desired phenomena can be revealed, a multiscale modeling approach can be recommended.

The classical models of continuum mechanics, which model a material as a continuous mass, have shown clear failures when applied for multiscale materials. Indeed, the classical mechanics cannot account for the discrete nature (hierarchical structure) of the material. Thus, many microstructural features, *e.g.* micro-slipping and twinning, cannot be obtained via the classical mechanics. Furthermore, the classical continuum mechanics lacks measures and/or length-scales which are needed to capture the nontraditional phenomena of nanomaterials, *e.g.* the grain size effects on the mechanics of nanocrystalline materials [1-3]. Therefore, multiscale modeling has been introduced to replace the classical mechanics for multiscale materials. Different reviews and discussions on multiscale materials modeling can be found in [4,5].

The existing models for multiscale materials follow two main approaches: hierarchical approach and hybrid approach. In the hierarchical approach [6-7] models for the different scales are run separately and some coupling parameters are utilized to couple these independently working models. In the hybrid approach [9-12], on the other hand, models are run concurrently over different spatial regions of a simulation [5].

Models for multiscale materials range from atomistic models to continuum mechanics-based models. Atomistic and molecular dynamics models represent the material as an assembly of atoms or molecules. The classical atomistic models depend on the classical interatomic potentials. However, some nanoscale phenomena were revealed only by utilizing newly developed interatomic interactions which depend on quantum-molecular mechanics hybrid models [12]. At larger scales ($0.1\text{-}10\mu m$), atomistic modeling of materials is computationally expensive, and the material is too small to be represented as a continuum. Therefore, mesoscale modeling has been developed to capture the phenomena at this intermediate scale. Mesoscale modeling techniques depend on merging redundant degrees of freedom by, for example, grouping atoms into particles (this is known as 'coarse graining') [5].

Some other models depend on the homogenization theory; these models can be used to reveal the macroscale phenomena of multiscale materials [4]. In these models, in order to model the hierarchical structure of the material, representative volume elements (RVEs) are introduced for the different scales. At the transition from one scale to the next upper-scale, a conventional micromechanics that depends on a RVE-averaging technique is applied. This method, however, shows discrepancies when applied for non-uniform microstructures and/or when the consideration of the deformability of the RVEs is a crucial [4,13].

An alternative technique is to model multiscale materials via microcontinuum models, *e.g.* multi-scale micromorphic theory [13]. These models remedy the drawbacks of the classical continuum mechanics for multiscale materials by the incorporation of new measures and length scales. The micromorphic theory [14,15] can emulate the discrete nature of the material microstructure by defining sub-continua (these continua may rigidly move and/or deform) for the different scales. However, the constitutive equations in the context of these models depend on additional material coefficients for which no correlations with the microstructure of the material at the different scales were defined. To remedy this drawback, Vernerey et al., [13] integrated the micromorphic theory with the homogenization theory to identify the material coefficients at the different scales.

Recently, new materials with unique wave propagation characteristics are designed. Periodic materials, metamaterials, and phononic materials are designed to provide exceptional properties such as band gaps [16-17], response directionality [19-21], negative index [22], and negative acoustic refraction [23]. The study of the wave propagation in these newly developed materials is of a particular interest. The manipulation of acoustic and optical waves in these materials is a fundamental problem with many potential applications including communication technologies, signal filtration and processing, and sensing applications [24]. Some of the aforementioned techniques of multiscale materials modeling were utilized to investigate the wave propagation characteristics of multiscale materials such as phononic materials [25] and metamaterials [26, 27].

In the present study, a new micromorphic model for multiscale materials is proposed. In the context of this model, the microstructure of a multiscale material is modeled consisting of a large number of deformable material particles. Moreover, the material coefficients of the proposed model are related to the material microstructure. First, the classical micromorphic model [14,15] in which the material particle is modeled exhibiting 12 degrees of freedom (DOFs), namely, 3-displacments, 3-rotations, and 6-microstrains is reconsidered. Thus, a 2-D model of a material confined between two particles is proposed to explain the contribution of each one of the 12-DOFs to the deformation energy of multiscale materials. The various deformation patterns in the confined material due to the 12-DOFs are depicted. It is demonstrated that some of these deformation patterns along with some degrees of freedom of the material particle can be eliminated. This process results in a reduced micromorphic model that can effectively model the mechanics of multiscale materials. Unlike the classical micromorphic model (which depends on 18 material coefficients for linear elastic materials), the reduced micromorphic model only depends on 8 material coefficients. This reduced form of the micromorphic model permits deriving relations between the material coefficients and the material microstructure.

Afterwards, the reduced micromorphic model is harnessed to investigate the wave propagation characteristics of multiscale materials. To this end, the wave dispersion relations according to the proposed model are derived. It is demonstrated that the proposed model can reflect 9 dispersion curves including three acoustic and six optical dispersion curves for a multiscale material. To demonstrate the effectiveness of the proposed micromorphic model, a case study of a phononic material is then carried out where its bandgap characteristics are investigated.

## 2. A Reduced Micromorphic Model for Multiscale Materials

The microstructure of a multiscale material is modeled consisting of a large number of building units (material particles) (*e.g.* grains, nanocrystals, nanoparticles, or unit-cells). Each of the building particles exhibits different types of degrees of freedom (DOFs), as shown in Figure 1. The governing equations of the deformation of multiscale materials are derived accounting for the microstrains of these material particles.

In fact, the particles may rigidly move (*e.g.* translate and/or rotate) and/or deform inside the material structure. Thus, generally, a particle may exhibit 12 DOFs (*i.e.* 3-displacements ($u_i$), 3-micro-rotations ($\vartheta_i$), and 6-microstrains ($s_{ij} = s_{ji}$)). The micromorphic theory [14,15,28] can account for all the 12-DOFs. Therefore, for a linear isotropic material, the micromorphic model depends on 18 material coefficients. This significantly limits its applicability because of the mathematical complications. Furthermore, the relations between these coefficients and the material's macro/micro-stiffnesses are difficult to obtain. In this study, an effective reduced form of the classical micromorphic model for multiscale materials is developed. With careful inspections of the 12 DOFs of the material particle, it is demonstrated that the contributions of some DOFs in comparison to the others are negligible. Moreover, some microstructural deformation patterns can be neglected. Therefore, omitting these less-affecting DOFs and deformation patterns, the micromorphic model can be simplified.

To this end, a 2-D model of two material particles and a confined (in-between) material is proposed, as shown in Figure 1. This model is considered to mimic, for example, two grains and the in-between grain boundary in a polycrystalline material, two unit cells and the in-between interatomic interaction in a single crystal, or two particles and the in-between matrix material in a composite. In this model, the two material particles undergo rigid displacements ($u_x$ and $u_y$) (Figure 1(a)), rigid rotations ($\vartheta_x$ and $\vartheta_z$) (Figure 2(b)), and in-plane strains ($s_{xx}$, $s_{yy}$, and $s_{xy}$) (Figure 1(c)). Depending on the DOFs of the two material particles, the confined material exhibits different deformation patterns. When the particles undergo rigid displacements, the confined material is stretched (Figure 1(a)(i)), sheared (Figure 1(a)(iii)), or rigidly moved (Figures 1(a)(ii) and 1(a)(iv)). On the other hand, the confined material may subject to twisting (Figure 1(b)(i)), axial bending (Figures 1(b)(iii) and 1(b)(iv)), or axial shaft-like-rotation (Figure 1(b)-(ii)) due to particles' rotations. In addition, as shown in Figure 1(c), the confined material deforms as the material

M. Shaat, arXiv [physics.app-ph], December 1, 2017.

particles deform. Thus, depending on the micro-strains of the material particle, the confined material is subjected to compression (Figure 1(c)(i)), rigid motion (Figure 1(c)(ii)), lateral stretching (Figure 1(c)(iv)), combined compression-stretch in the lateral direction (Figure 1(c)(iii)), or axial bending (Figures 1(c)(iiv) and 1(c)(iiiv)). This proposed model presents a detailed explanation for the DOFs of the material particle and their contributions to multiscale materials deformation. Table 1 summarizes the 12-DOFs and their contributions to the deformation of a confined material with $x$−axis is its longitudinal axis.

Table 1: Deformation patterns of a confined material between two deformable particles (see Figure 1).

| DOF Type | Number of DOFs | DOF | Deformation Pattern |
|---|---|---|---|
| Displacements ($u_i$) | 3 | $u_x$ | Stretching |
| | | $u_y$ and $u_z$ | Shearing |
| Rotations ($\vartheta_i$) | 3 | $\vartheta_x$ | Twisting |
| | | $\vartheta_y$ and $\vartheta_z$ | Axial bending |
| Micro-strains ($s_{ij}$) | 6 | $s_{xx}$ | Compression/stretching |
| | | $s_{yy}$ and $s_{zz}$ | Lateral stretching/compression |
| | | $s_{xy}$, $s_{xz}$, and $s_{yz}$ | Axial bending |

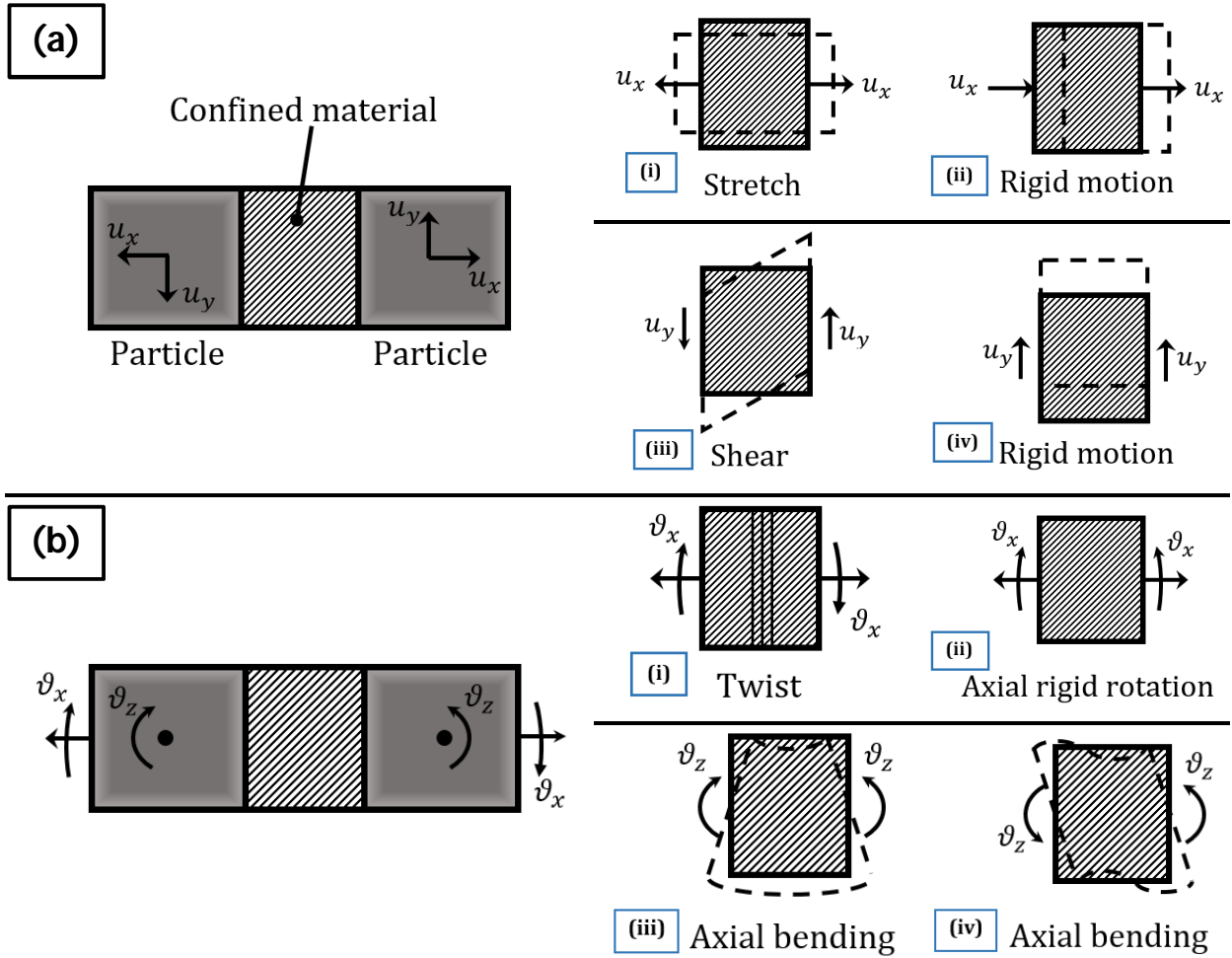

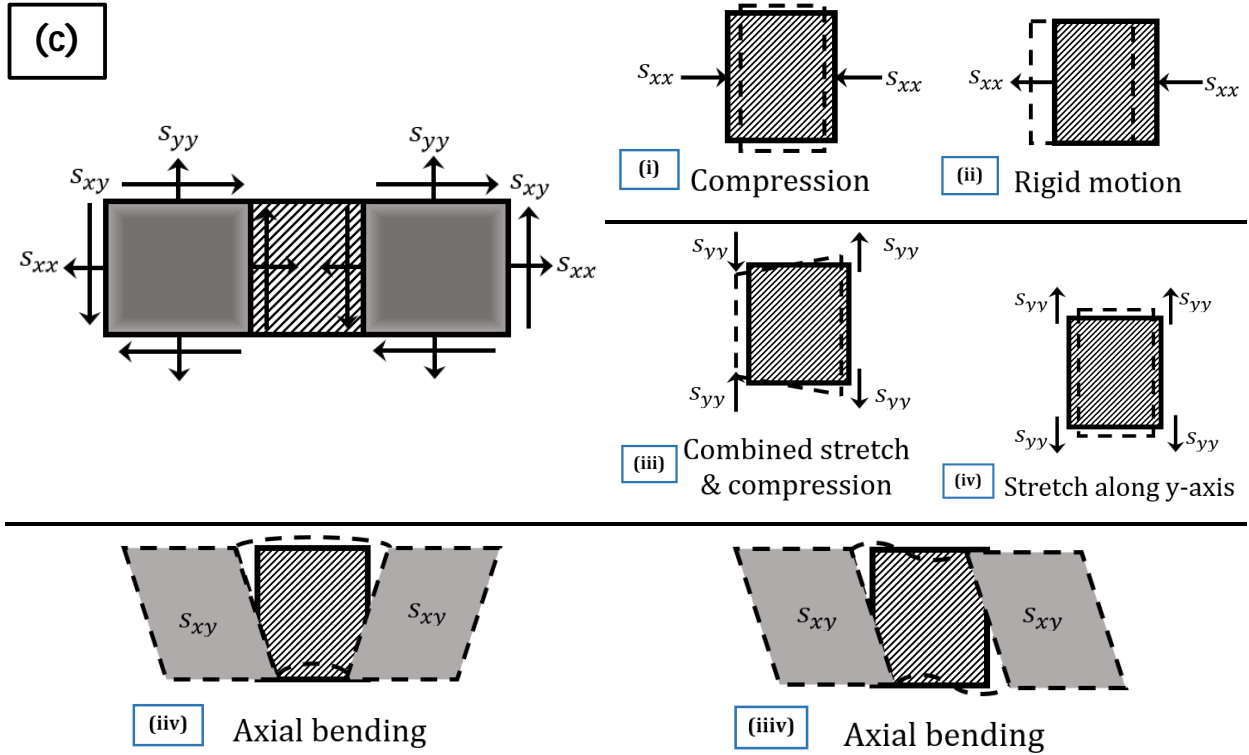

Figure 1: 2-D model illustrating the deformation patterns of a material confined between two particles in a multiscale material. The deformation patterns in the confined material due to (a) rigid displacements, (b) rigid rotations, and (c) microstrains of the particles are depicted.

Indeed, some of the deformation patterns of the confined material can be eliminated. First of all, by definition, the rigid-body motions (Figures 1(a)(ii) and 1(a)(iv), 1(b)(ii), and 1(c)(ii)) should be omitted since they do not contribute any deformation in the material. Second, the stretch and/or compression in the confined material due to normal microstrains (Figures 1(c)(i), 1(c)(iii), and 1(c)(iv)) are, in fact, negligible in comparison to the stretch/compression/shear due to the displacements of the particles (Figures 1(a)(i) and 1(a)(iii)). Furthermore, the axial twisting in the confined material (Figure 1(b)(i)) is negligible comparing to the axial bending (Figures 1(b)(iii) and 1(b)(iv)). Therefore, to a great extent, it is acceptable to eliminate these less-affecting deformation patterns (Figures 1(c)(i), 1(c)(iii), 1(c)(iv), and Figure 1(b)(i)). Finally, as shown in Figures 1(b) and 1(c), the confined material experiences the same deformation patterns (*i.e.* axial bending) when the two particles exhibit rotations or shear microstrains. Indeed, the axial bending due to one of the two types of DOFs can be approximately compensated by adjusting the materials coefficients in the constitutive equations of the other DOF type. Therefore, a further simplification can be achieved by modeling the axial bending in the confined material only via the microstrains and approximately compensating the axial bending due to the rigid rotations of the material particle in the constitutive equations.

By eliminating the aforementioned deformation patterns, the following kinematical variables can be defined to account for the deformation of multiscale materials:

$$\varepsilon_{ij} = \varepsilon_{ji} = \frac{1}{2}(u_{i,j} + u_{j,i}) \tag{1}$$

$$\chi_{ijk} = \chi_{ikj} = s_{jk,i}, \text{ i.e. } \chi_{ijj} = 0 \tag{2}$$

$$\gamma_{ij} = \gamma_{ji} = \varepsilon_{ij} - s_{ij} \tag{3}$$

where $\varepsilon_{ij}$ is the conventional strain dyadic of the classical theory which accounts for the stretch and shear in the material due to the material particle's displacements, $u_i$. $\chi_{i(jk)}$ is a triadic, that is symmetric on the last two indexes, introduced as the gradient of the microstrain dyadic, $s_{ij}$. Because the stretch and compression in the material due to normal microstrains are eliminated, the $\chi_{ijj}$ components are omitted. Because they are independent, $\gamma_{ij}$ is introduced to account for the coupling between the microstrain ($s_{ij}$) and the macro-strain ($\varepsilon_{ij}$).

Utilizing the kinematical variables in equations (1)-(3), the deformation energy density function of multiscale materials can be defined as follows:

$$\overline{W}(x) = \frac{1}{2}\lambda_m s_{ii}s_{jj} + \mu_m s_{ij}s_{ij} + \frac{1}{2}\lambda\gamma_{ii}\gamma_{jj} + \mu\gamma_{ij}\gamma_{ij} + \lambda_c\gamma_{ii}s_{jj} + 2\mu_c\gamma_{ij}s_{ij} \\ + \frac{1}{2}\lambda_m\ell_1^2(\chi_{iik}\chi_{jjk} + \chi_{ijk}\chi_{jik}) + \mu_m\ell_2^2\chi_{ijk}\chi_{ijk} \tag{4}$$

where $\lambda_m$ and $\mu_m$ are the elastic moduli of the microstructure (micro-moduli). $\lambda$ and $\mu$ denote the elastic moduli of the confined material between two particles (see Figure 1). For polycrystalline materials, for example, $\lambda_m$ and $\mu_m$ denote the elastic moduli of the grains while $\lambda$ and $\mu$ are the elastic moduli of the grain boundary. These elastic moduli can be determined using micro-tensile tests for nano-sized specimens contain two grains. $\lambda_c$ and $\mu_c$ are two elastic moduli account for the coupling between the microstrain and the macro-strain. $\ell_1$ and $\ell_2$ are length scale parameters that can be related to the material particle size (*e.g.* grain size, lattice parameter, etc).

The deformation energy function as introduced in equation (4) depends on material coefficients that are related to the material's microstructure. Thus, this equation can be effectively used to determine the deformation energy of multiscale materials.

According to equation (4), the constitutive equations can be derived in the form:

$$t_{ij} = t_{ji} = \frac{\partial \overline{W}}{\partial s_{ij}} = \lambda_m s_{ll}\delta_{ij} + 2\mu_m s_{ij} + \lambda_c\gamma_{ll}\delta_{ij} + 2\mu_c\gamma_{ij} \tag{5}$$

$$\tau_{ij} = \frac{\partial \overline{W}}{\partial \gamma_{ij}} = \lambda \gamma_{ll} \delta_{ij} + 2\mu \gamma_{ij} + \lambda_c s_{ll} \delta_{ij} + 2\mu_c s_{ij} \tag{6}$$

$$m_{ijk} = \frac{\partial \overline{W}}{\partial \chi_{ijk}} = \lambda_m \ell_1^2 \chi_{mmk} \delta_{ij} + \mu_m \ell_2^2 \chi_{ijk} + \lambda_m \ell_1^2 \chi_{jik} \tag{7}$$

where $t_{ij}$ is a micro-stress tensor which is a symmetric tensor. $\tau_{ij}$ is introduced as a stress tensor that measures the residual stresses in the macro-material due to the difference in microstrain and macrostrain fields. For rigid particles (zero micro-strains), this tensor reduces to the conventional force-stress tensor of the classical elasticity. $m_{ijk}$ is a triadic that is conjugate to $\chi_{ijk}$. $\tau_{ij}$ and $m_{ijk}$ are presented to capture the effects of microstrains on the macro-scale deformation of multiscale materials.

To derive the equations of motion, Hamilton's principle is applied. Therefore, the first variations of the kinetic energy $\delta K$, the stored energy $\delta U$, and the work-done $\delta Q$ are expressed as follows (considering a continuum occupies a volume $V$ and a surface $S$):

$$\delta K = -\int_V \left( \rho \ddot{u}_i \delta u_i + \rho_m J \ddot{s}_{jk} \delta s_{jk} \right) dV \tag{8}$$

$$\delta U = \delta \int_V \overline{W} \, dV = \int_V \left( -(\tau_{ji,j}) \delta u_i - (\tau_{jk} - t_{jk} + m_{ijk,i}) \delta s_{jk} \right) dV \\ + \int_S \left( (n_j \tau_{ji}) \delta u_i + (n_i m_{ijk}) \delta s_{jk} \right) dS \tag{9}$$

$$\delta Q = \int_V \left( f_i \delta u_i + H_{jk} \delta s_{jk} \right) dV + \int_S \left( \bar{t}_i \delta u_i + \bar{m}_{jk} \delta s_{jk} \right) dS \tag{10}$$

where $\rho$ represents the mass density of the macro-scale material (confined material). $\rho_m$ is the mass density of the material particle, and $J$ denotes its micro-inertia. $f_i$ and $H_{jk}$ are, respectively, body forces and body higher-order-moments.

Therefore, the equations of motion are obtained in the form:

$$\tau_{ji,j} + f_i = \rho \ddot{u}_i \tag{11}$$

$$m_{ijk,i} + \tau_{jk} - t_{jk} + H_{jk} = \rho_m J \ddot{s}_{jk} \tag{12}$$

and the corresponding natural boundary conditions are given by:

$$n_j \tau_{ji} = \bar{t}_i \tag{13}$$

$$n_i m_{ijk} = \bar{m}_{jk} \tag{14}$$

Equations (11) and (12) represent nine equilibrium equations that govern the micro-deformations of multiscale materials.

It should be mentioned that the proposed model as presented in equations (11) and (12) can be directly applied for two-scale materials with uniform microstructures. In order to generalize the proposed model for multiple-scale materials, the model can be sequentially applied starting from the lower scale following the hierarchical structure of the material. Moreover, the proposed model can be considered as a simplification of the multi-scale micromorphic model proposed by Vernerey et al. [13]. Thus, for non-uniform/heterogeneous microstructures, the proposed model can be integrated with a homogenization approach following the multi-scale micromorphic model [13,29]. This homogenization approach is used to identify the micro-moduli of the particle at the considered scale from the underlying microstructure of the particle.

Another form of the micromorphic model that is recently proposed by Neff et al. [30] is the relaxed micromorphic model. Unlike the proposed reduced micromorphic model, the relaxed micromorphic model depends on the curl of the micro-deformation tensor as a micro-dislocation measure. Moreover, the coupling between the micro-strain and the macro-strain is eliminated in the relaxed model. Recently, the relaxed micromorphic model was utilized to model the wave propagation in metamaterials [27, 31-33].

Examples of materials that can be modeled via the proposed microcontinuum model are shown in Figure 2. For two-scale materials with uniform microstructures, a single RVE can be defined to represent the whole microstructure. For instance, Figure 2(a) shows TEM micrograph and SAED pattern of ultra-fine grained titanium processed by equal channel angular pressing [34]. The microstructure is homogeneous (diffraction spots form circles in the SAED pattern) with nanograins (as the building blocks) are equiaxed with different crystallographic orientations. This type of materials can be considered as a two-scale material and hence the proposed model can be directly applied. The micro-moduli ($\lambda_m$ and $\mu_m$) are defined for the grains while the grain boundaries are considered as confined materials with $\lambda$ and $\mu$ as elastic moduli.

Figure 2(b) shows a case of a three-scale material for which two RVEs should be defined. The figure shows a superparticle that is assembled from interacting-artificial atoms [35]. As shown in the TEM micrograph, atoms are well arranged forming a superparticle with a well-defined shape. Moreover, high uniformity in the spaces between the diffraction spots in the SAED pattern can be observed indicating a uniform-homogeneous structure of the superparticle. For this superparticle, a representative cell (or a micro-RVE) that is composed of many atoms is defined. Within this micro-RVE, atoms are considered rigidly moving where the proposed model is applied with $s_{ij} = 0$ and $\lambda$ and $\mu$ defined according to the interatomic potential. Then, the proposed model is applied again where the obtained deformation fields of the micro-RVE are used to form the microstrain field and its corresponding moduli for a macro-scale RVE.

The TEM micrographs and SAED patterns of graphene-ZnS composite [36] and a monolayer assembled from nanocubes [37] are, respectively, shown in Figures 2(c) and 2(d). The TEM micrographs show ZnS nanoparticles are uniformly dispersed in the matrix, and the monolayer possesses a uniform structure (*i.e.* numerous diffraction spots which are arranged in circles can be seen in SAED patterns indicating highly uniform structures). A homogenization RVE can be defined for these materials to identify the micro-moduli $\lambda_m$ and $\mu_m$. Because of the incorporated homogenization, the microstructure converts to be a one-phase-homogeneous microstructure. This permits considering $\lambda = \lambda_m$ and $\mu = \mu_m$ when applying the reduced micromorphic model.

Next, the reduced micromorphic model is utilized to investigate the wave propagation in multiscale materials. Two case studies of phononic materials are considered to show the effectiveness of the proposed model.

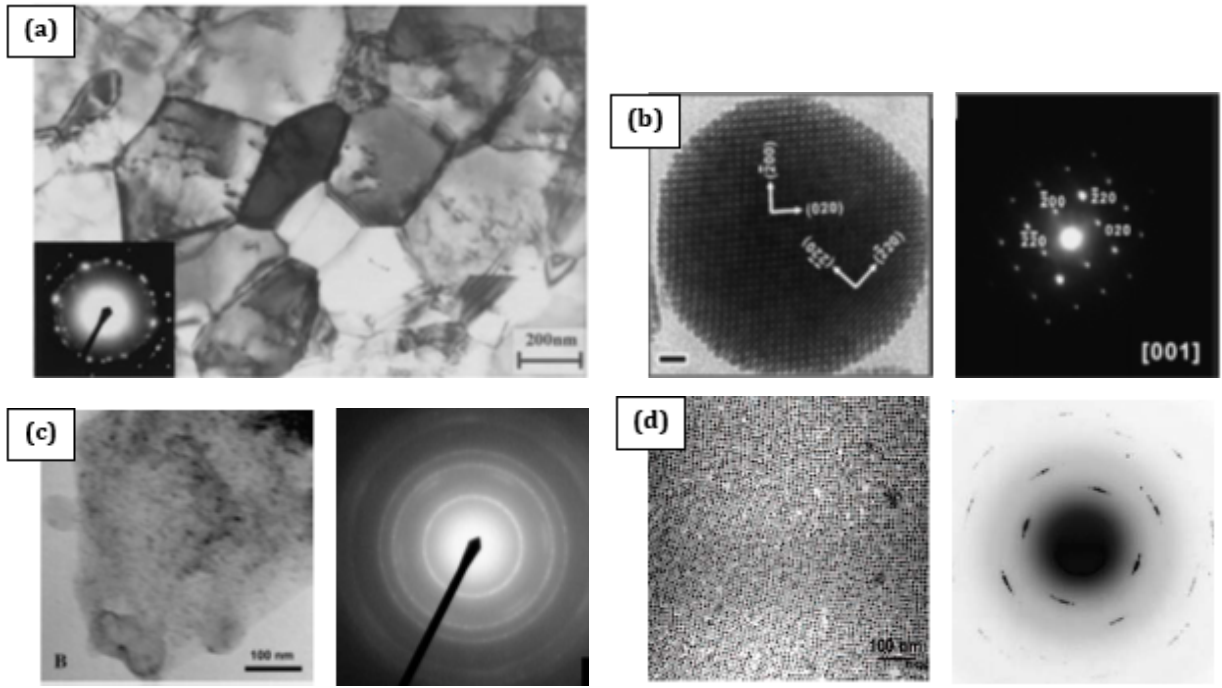

Figure 2 TEM micrographs and SAED patterns of (a) titanium processed by equal channel angular pressing [34], (b) a superparticle obtained as a supercrystalline collection of artificial atoms [35], (c) graphene-ZnS composite [36], (d) a monolayer assembled from nanocubes [37].

## 3. Application to Wave Propagation in Multiscale Materials

To form the wave dispersion relations in a multiscale material, the reduced micromorphic model is utilized. The following harmonic waves along the $x$-direction are assumed:

$$u_i = iU_i e^{i(kx-\omega t)} \tag{15}$$

$$s_{ij} = S_{ij} e^{i(kx-\omega t)} \tag{16}$$

where $k$ denotes the wave number and $\omega$ is the wave frequency.

These harmonic functions represent the considered nine-DOFs of a material particle. $u_i$ are three-displacements while $s_{ij}$ denote the six-microstrains.

To form the wave dispersions, first, the kinematical variables, $\varepsilon_{ij}$, $\gamma_{ij}$, and $\chi_{ijk}$, are formed by substituting the harmonic functions (equations (15) and (16)) into equations (1)-(3). Then, the constitutive relations are formed utilizing the obtained kinematical variable and according to equations (5)-(7). The nine-equations of motions can be then formed according to equations (11) and (12). By dropping the body forces and the body moments, equations of motions can be rearranged as follows:

$$\left[\left((2\mu+\lambda)k^2 - \rho\omega^2\right)U_x + (2\mu - 2\mu_c)kS^D + (2\mu + 3\lambda - 2\mu_c - 3\lambda_c)k\hat{S}\right]e^{i(kx-\omega t)} = 0 \tag{17(a)}$$

$$\left[(2\mu - 2\mu_c)kU_x + \left(\left(\frac{3}{2}\mu_m\ell_2^2 + 2\lambda_m\ell_1^2\right)k^2 + 3\mu + 3\mu_m - 6\mu_c - \frac{3}{2}\rho_m J\omega^2\right)S^D \right.$$
$$\left. + 2\lambda_m\ell_1^2 k^2 \hat{S}\right]e^{i(kx-\omega t)} = 0 \tag{17(b)}$$

$$\left[(2\mu + 3\lambda - 2\mu_c - 3\lambda_c)kU_x + 2\lambda_m\ell_1^2 k^2 S^D \right.$$
$$\left. + \left((3\mu_m\ell_2^2 + 2\lambda_m\ell_1^2)k^2 + (9\lambda + 6\mu + 9\lambda_m + 6\mu_m - 18\lambda_c - 12\mu_c)\right.\right.$$
$$\left.\left. - 3\rho_m J\omega^2\right)\hat{S}\right]e^{i(kx-\omega t)} = 0 \tag{17(c)}$$

$$\left[(\mu k^2 - \rho\omega^2)U_y + (2\mu - 2\mu_c)kS_{xy}\right]e^{i(kx-\omega t)} = 0 \tag{17(d)}$$

$$\left[(2\mu - 2\mu_c)kU_y + \left((2\mu_m\ell_2^2 + 2\lambda_m\ell_1^2)k^2 + 4\mu + 4\mu_m - 8\mu_c - 2\rho_m J\omega^2\right)S_{xy}\right]e^{i(kx-\omega t)} = 0 \tag{17(e)}$$

$$\left[(\mu k^2 - \rho\omega^2)U_z + (2\mu - 2\mu_c)kS_{xz}\right]e^{i(kx-\omega t)} = 0 \tag{17(f)}$$

$$\left[(2\mu - 2\mu_c)kU_z + \left((2\mu_m\ell_2^2 + 2\lambda_m\ell_1^2)k^2 + 4\mu + 4\mu_m - 8\mu_c - 2\rho_m J\omega^2\right)S_{xz}\right]e^{i(kx-\omega t)} = 0 \tag{17(g)}$$

$$\left(\mu_m\ell_2^2 k^2 + \frac{2}{3}(3\mu + 3\mu_m - 6\mu_c) - \rho_m J\omega^2\right)S_{yz}e^{i(kx-\omega t)} = 0 \tag{17(h)}$$

$$\left(\mu_m\ell_2^2 k^2 + \frac{2}{3}(3\mu + 3\mu_m - 6\mu_c) - \rho_m J\omega^2\right)(S_{yy} - S_{zz})e^{i(kx-\omega t)} = 0 \tag{17(I)}$$

where $\hat{S} = \frac{1}{3}(S_{xx} + S_{yy} + S_{zz})$ and $S^D = S_{xx} - \hat{S}$.

It is clear that 9-equations of motion are obtained, as shown in Equation (17). The dispersion relations according to the reduced micromorphic model can be then obtained as follows:

(1) Longitudinal acoustic (LA), Longitudinal optic (LO), and Longitudinal dilatational optic (LDO) waves:

$$\begin{vmatrix} A_{11}k^2 - \rho\omega^2 & A_{12}k & A_{13}k \\ A_{12}k & A_{22}k^2 + B_{22} - \frac{3}{2}\rho_m J\omega^2 & A_{23}k^2 \\ A_{13}k & A_{23}k^2 & A_{33}k^2 + B_{33} - 3\rho_m J\omega^2 \end{vmatrix} = 0 \tag{18}$$

(2) Transverse acoustic (TA) and Transverse optic (TO) waves:

$$\begin{vmatrix} \bar{A}_{11}k^2 - \rho\omega^2 & A_{12}k \\ A_{12}k & \bar{A}_{22}k^2 + \bar{B}_{22} - 2\rho_m J\omega^2 \end{vmatrix} = 0 \tag{19}$$

(3) Transverse shear optic (TSO) waves:

$$A_{44}k^2 + \frac{2}{3}B_{22} - \rho_m J\omega^2 = 0 \tag{20}$$

The coefficients presented in equations (18)-(20) are obtained as follows:

$$\begin{aligned}
A_{11} &= 2\mu + \lambda \\
A_{12} &= 2\mu - 2\mu_c \\
A_{13} &= 2\mu + 3\lambda - 2\mu_c - 3\lambda_c \\
A_{22} &= \frac{3}{2}\mu_m \ell_2^2 + 2\lambda_m \ell_1^2 \\
A_{23} &= 2\lambda_m \ell_1^2 \\
A_{33} &= 3\mu_m \ell_2^2 + 2\lambda_m \ell_1^2 \\
A_{44} &= \mu_m \ell_2^2 \\
\bar{A}_{11} &= \mu \\
\bar{A}_{22} &= 2\mu_m \ell_2^2 + 2\lambda_m \ell_1^2 \\
B_{22} &= 3\mu + 3\mu_m - 6\mu_c \\
\bar{B}_{22} &= 4\mu + 4\mu_m - 8\mu_c \\
B_{33} &= 9\lambda + 6\mu + 9\lambda_m + 6\mu_m - 18\lambda_c - 12\mu_c
\end{aligned} \tag{21}$$

According to equations (18)-(20), the reduced micromorphic model can reflect nine-wave dispersions for multiscale materials.

## 4. Wave Propagation in Phononic Materials

Phononic crystals (materials) are artificial materials consisting of periodic structures of inclusions embedded in a matrix. Controlling their microstructures, phononic materials can be designed with the capability of isolating wave propagation frequencies within a specific range of frequencies. The isolated region known as the bandgap [24]. The elastic band structure of phononic materials was investigated by Sigalas and Economou [38-40] and Kushwaha et al. [41,42]. It was demonstrated that phononic materials exhibit absolute phononic bandgaps depending on the material composition [24,42].

In the present study, to investigate the elastic band structure and the absolute bandgaps in phononic materials, the reduced micromorphic model is harnessed to model the wave propagation in a phononic material type. A schematic of the structure of the considered phononic material is shown in Figure 3. This material is considered with a periodic microstructure with hard inclusions embedded in a soft matrix. A square lattice (unit cell) is considered to represent the microstructure, as shown in Figure 3(b).

In this study, the phononic material is considered with two different material-composition. Thus, two approaches are employed for the considered two cases of the phononic material. In one approach, the original heterogeneous microstructure is treated by the reduced micromorphic model. In the second approach, a homogenization model is incorporated to derive an equivalent homogeneous microstructure to be treated by the reduced micromorphic model.

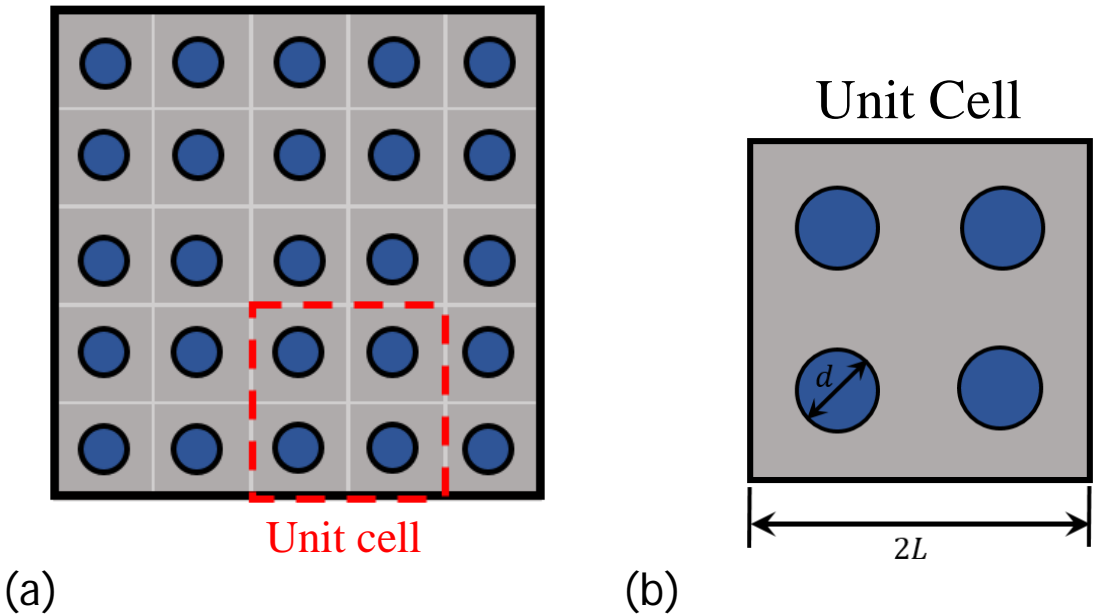

Figure 3: (a) A schematic of the considered phononic material with aluminum matrix and stiff inclusions. The unit cell is presented by red dashed square. The schematic of the unit cell is shown in (b).

## 4.1 Case 1

In the first material example, the phononic material is composed of an aluminum matrix (with $\lambda = 40.60\ GPa$, $\mu = 27.31\ GPa$, and $\rho = 2700\ kg/m^3$) in which ten-times stiffer and heavier inclusions are embedded. According to the reduced micromorphic model, the inclusions are considered exhibiting microstrains as well as displacements while the aluminum matrix is considered as the confined material. Therefore, the original elastic moduli of the inclusions are defined as the micro-moduli $\lambda_m$ and $\mu_m$ while the elastic moduli of the confined material ($\lambda$ and $\mu$) are considered equal to the ones of the aluminum matrix. For this case, the coupling moduli ($\lambda_c$ and $\mu_c$) should be employed to capture the inclusion size effects. For this reason, the coupling moduli are related to the elastic moduli of an equivalent homogeneous unit cell. The equivalent elastic moduli ($\lambda_e$ and $\mu_e$) of the homogeneous unit cell are determined using the rule of mixture as follows:

$$\begin{aligned}\lambda_e &= f\lambda_m + (1-f)\lambda \\ \mu_e &= f\mu_m + (1-f)\mu \\ \rho_e &= f\rho_m + (1-f)\rho\end{aligned} \quad (22)$$

where $f = \frac{\pi}{4}\left(\frac{d}{L}\right)^2$ is the volume fraction of the inclusions. Table 2 shows the material properties of the considered phononic material and their corresponding parameters of the reduced micromorphic model.

It should be mentioned that the coupling moduli ($\lambda_c$ and $\mu_c$) were completely omitted in the relaxed micromorphic model [27, 30-33]. In the proposed model, in contrast, the coupling moduli are employed to account for the inclusion size effects.

**Table 2**: Material characteristics of the considered phononic material and their corresponding parameters in the context of the reduced micromorphic model.

| **Phononic material properties** | |
|---|---|
| Matrix: $\lambda = 40.60 GPa$, $\mu = 27.31\ GPa$, and $\rho = 2700\ kg/m^3$ | |
| Inclusions: $\lambda = 406\ GPa$, $\mu = 273.1\ GPa$, and $\rho = 27000\ kg/m^3$ | |
| **Parameters of the reduced micromorphic model** | |
| Unit cell width | $2L = 2\ m$ |
| Micro-inertia density | $J = (2L)^2/32 = 0.125\ m^2$ |
| Maximum wave number | $k_{max} = 2\pi/2L = \pi\ m^{-1}$ |
| Micro-moduli | $\lambda_m = 406\ GPa$ and $\mu_m = 273.1\ GPa$ |
| Macro-moduli | $\lambda = 40.60\ GPa$ and $\mu = 27.31\ GPa$ |
| Mass densities | $\rho_m = 27000\ kg/m^3$ and $\rho = 2700\ kg/m^3$ |
| Coupling moduli | $\lambda_c = -0.9\lambda_e$ and $\mu_c = -0.9\mu_e$ |
| Length scales | $\ell_1^2 = 0$ and $\ell_2^2 = 0$ |

The proposed model is utilized to plot the wave dispersions of the considered phononic material for different $d/L$ ratios (where $d$ is the inclusion's diameter and $2L$ is the unit cell length). It should be mentioned that the same case study was previously considered by Casadei [25,43] utilizing a Generalized Multiscale Finite Element Method (GMsFEM). Therefore, the wave propagation characteristics of the considered phononic material as obtained by the reduced micromorphic model are presented in parallel to those obtained using the GMsFEM, as shown in Figure 4. As depicted in the figure, the proposed model effectively gives an excellent match with the GMsFEM. This demonstrates the effectiveness of the proposed model to model the wave propagation in multiscale materials.

By inspecting Figure 4, it is clear that the increase in the inclusion size is accompanied with an increase in the bandgap width. This behavior can be attributed to the fact that the bandgap behavior of phononic materials depends on the contrast between the constituent materials and the lattice structure. Thus, the increase in the inclusion size increases the contrast which results in an increase in the bandgap. Moreover, the considered lattice structure (square unit cell) gives a continuous increase in the bandgap width with the increase in the inclusion filling factor (volume fraction) [24]. These bandgap behaviors of phononic materials are effectively emulated by the proposed model. Thus, the coupling moduli are used as essential measures for the contrast between the micro-medium and the macro-medium. Indeed, the increase in the filling factor increases the difference between the micro-strain and the macro-strain, and hence increases the bandgap width.

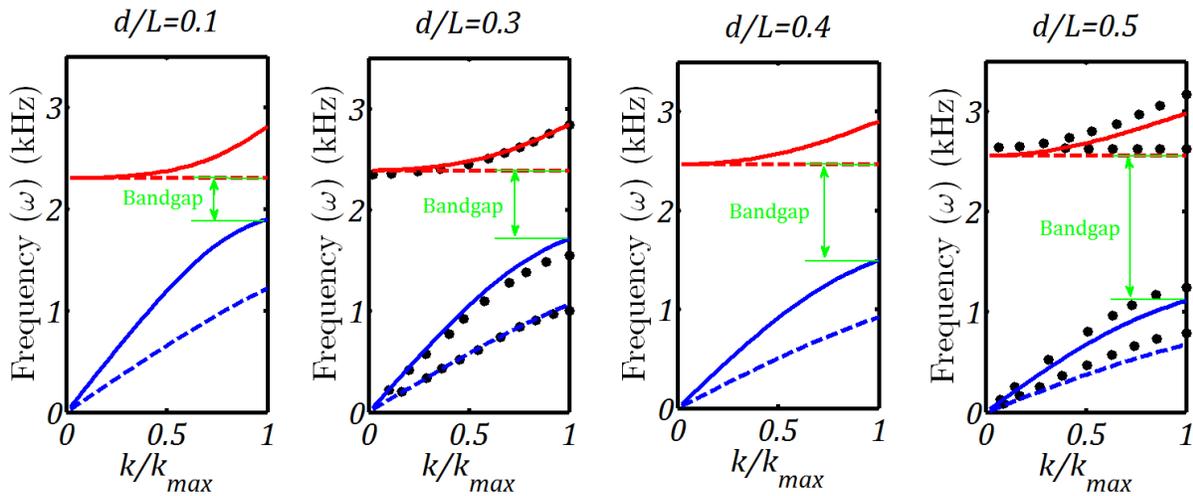

Figure 4: Dispersion curves of the considered phononic material for different inclusion sizes, $d/L = 0.1 \rightarrow 0.5$. The longitudinal acoustic and optical phonons are presented as solid lines while the transverse acoustic and optical phonons are presented by dashed lines. Dark circles are the results as obtained by Casadei [25,43] using the Generalized Multiscale Finite Element Method (GMsFEM).

The bandgap behavior of phononic materials can be mathematically described using the reduced micromorphic model. According to the derived wave dispersion relations in equations (18)-(20), the cut-off frequency of the longitudinal and transverse optical modes can be determined as follows:

$$\omega_c = \sqrt{2\mu + 2\mu_m - 4\mu_c} \qquad (23)$$

For the considered phononic material and the defined parameters of the reduced micromorphic model, the cut-off frequency becomes $\omega_c = \sqrt{2\mu + 2\mu_m + 3.6\mu_e}$. Thus, the increase in the filling factor increases the effective modulus, $\mu_e$, and hence increases the cut-off frequency. This can be clearly seen in Figure 4. Moreover, it can be observed that the increase in the filling factor is accompanied with a decrease in the slope of the acoustic curves. These two trends cause the bandgap width increase with the increase in the filling factor.

The depicted results in Figure 4 demonstrate the effectiveness of the reduced micromorphic model in revealing the wave propagation characteristics of phononic materials.

## 4.2 Case 2

In the second material example, the phononic material is considered consisting of an Epoxy matrix in which silicon inclusions are arranged as shown in Figure 3. Another approach is utilized to treat this phononic material type. First, the equivalent elastic moduli ($\lambda_e$ and $\mu_e$) of the composite structure are derived using the rule of mixture (equation (22)). Then, the micro-moduli ($\lambda_m$ and $\mu_m$) and the macro-moduli ($\lambda$ and $\mu$) are considered equal to the obtained equivalent elastic moduli. This approach permits eliminating the role of the coupling moduli resulting in a more reduction in the micromorphic model. Thus, the contribution of the inclusion filling factor and the contrast between the material constituents can be automatically emulated by the equivalent elastic moduli. The material properties along with the parameters of the reduced micromorphic model for the considered epoxy-silicon material are shown in Table 3.

Figure 5 shows the wave dispersions in a pure silicon (a unity filling factor $f = 1$). When compared with the experimental results, the reduced micromorphic model gives a good match. The reduced micromorphic model effectively matches the experimental results for the acoustic dispersion curves. Moreover, the same trends of the optical branches are obtained using the model with small discrepancies at high wave numbers.

Figure 6 shows the evolution of the bandgap width with the increase in the filling factor for the epoxy-silicon phononic material. The figure shows the forbidden frequencies (gray area) for the different filling factor values. Moreover, the frequencies that define the lower and higher boundaries of the bandgaps are depicted by solid lines. The lower boundary-frequencies are acoustic frequencies while the upper boundary frequencies are optical frequencies. As shown in the figure, the increase in the filling factor is accompanied with an increase in the bandgap width. Thus, the maximum bandgap is achieved at $f = 1$. This results match the observations of the bandgap structure of square unit cells [24].

**Table 3**: Material characteristics of the considered phononic material (case 2) and their corresponding parameters in the context of the reduced micromorphic model.

**Phononic material properties**

Matrix (epoxy): $\lambda = 4.43\,GPa$, $\mu = 1.59\,GPa$, and $\rho = 1180\,kg/m^3$
Inclusions (silicon): $\lambda = 63.9\,GPa$, $\mu = 50.9\,GPa$, and $\rho = 2331\,kg/m^3$

**Parameters of the reduced micromorphic model**

| | |
|---|---|
| Lattice constant | $a = 0.543\,nm$ |
| Micro-inertia density | $J = a^2/32 = 0.009214\,nm^2$ |
| Maximum wave number | $k_{max} = 2\pi/a = 11.57\,nm^{-1}$ |
| Micro-moduli | $\lambda_m = \lambda_e$ and $\mu_m = \mu_e$ |
| Macro-moduli | $\lambda = \lambda_e$ and $\mu = \mu_e$ |
| Mass densities | $\rho_m = \rho_e$ and $\rho = \rho_e$ |
| Coupling moduli | $\lambda_c = 0$ and $\mu_c = 0$ |
| Length scales | $\ell_1^2 = 0.04a^2$ and $\ell_2^2 = -0.1a^2$ |

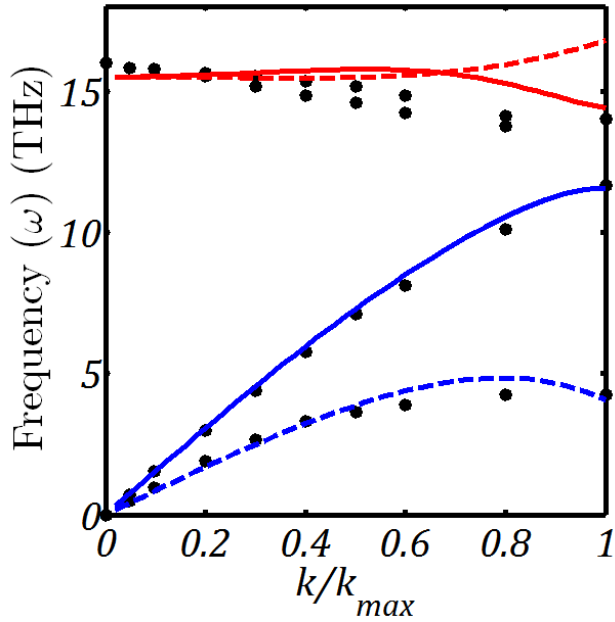

Figure 5: Dispersion curves of silicon as obtained by the reduced micromorphic model in comparison to the experimental-based curves [44]. The longitudinal acoustic and optical phonons are presented as solid lines while the transverse acoustic and optical phonons are presented by dashed lines. Dark circles are the experimental results.

Figure 6 shows a bandgap map for the considered phononic materials. In fact, these bandgap maps are needed in order to design phononic materials for various applications. Thus, depending on the prescribed performance, a composition of the phononic material can be recommended using these maps. It should be noted that the configuration of these bandgap maps depends on the lattice structure. For instance, Khelif and Adibi [24] obtained different bandgap maps for square, hexagonal, and honeycomb lattices.

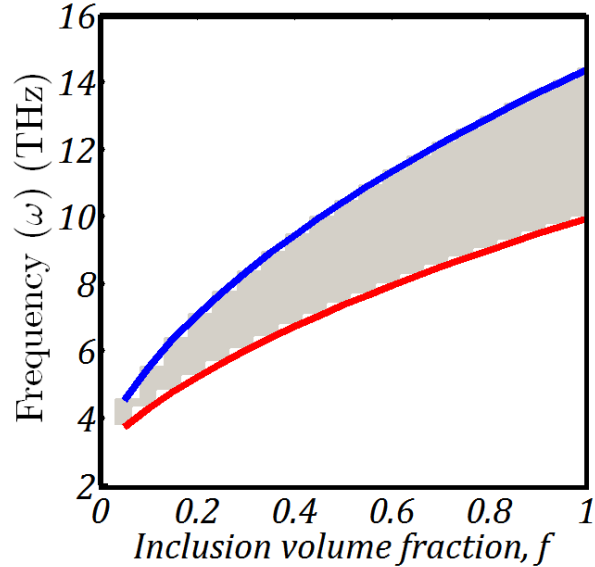

Figure 6: A bandgap map of the evolution of the absolute phononic bandgaps in an epoxy-silicon phononic material (with the epoxy is the matrix material in which silicon is embedded) with the increase in the silicon filling factor, $f$. Shaded region denotes the forbidden frequencies. The red solid line denotes the lower (acoustic) boundary-frequencies of bandgaps while the blue solid line is the upper (optical) boundary-frequencies of bandgaps.

## Conclusions

A 2-D model of a material confined between two particles was proposed to investigate the microstructural deformation patterns of multiscale materials due to internal microstrains and microrotations. It was revealed that some deformation patterns and degrees of freedom can be disregarded. This resulted in developing a new reduced micromorphic model for multiscale materials. This model depends on microstrains and macroscopic strain residuals to depict the deformation energy of multiscale materials.

The reduced micromorphic model was then utilized to investigate the wave propagation characteristics in multiscale materials. It was shown that this model can depict 9 dispersion curves for a two-scale material. After that, the model was harnessed to study the wave propagation characteristics, band structure, and bandgap evolution of phononic materials. Two examples of phononic material were considered in the performed analyses. It was demonstrated that phononic materials exhibit absolute phononic bandgaps depending on the lattice structure and the contrast between the constituent materials. Moreover, phononic materials with square lattices give a continuous increase in the bandgap width with the increase in the inclusion filling factor.

Finally, it was demonstrated that the reduced micromorphic model is an effective tool to determine the bandgap maps for multiscale materials such as phononic materials. These bandgap maps are needed in order to design phononic materials for various applications.